\documentclass[submission,copyright,creativecommons]{eptcs}

\usepackage{iftex}
\usepackage{amsmath}
\usepackage{graphicx}
\usepackage{multirow}
\usepackage{url} 
\usepackage{enumitem}
\usepackage{xcolor}
\usepackage{listings}
\lstdefinestyle{myPrologstyle}
{
    language=Prolog,
    basicstyle = \ttfamily\color{blue},
    moredelim = [s][\color{black}]{(}{)},
    literate =
        {:-}{{\textcolor{black}{:-}}}2
        {,}{{\textcolor{black}{,}}}1
        {.}{{\textcolor{black}{.}}}1
}

\usepackage{comment}

\ifpdf
  \usepackage{underscore}         
  \usepackage[T1]{fontenc}        
\else
  \usepackage{breakurl}           
\fi

\title{Subgraph Isomorphism: Prolog vs. Conventional}
\def\titlerunning{Subgraph Isomorphism: Prolog vs. Conventional}
\author{Claire Y. Yin
\institute{Computer Science and Engineering\\
University of Notre Dame\\
Notre Dame, IN USA 46556}
\and
Peter M. Kogge
\institute{Computer Science and Engineering\\
University of Notre Dame\\
Notre Dame, IN USA 46556}
}
\def\titlerunning{Subgraph Isomorphism: Prolog vs. Conventional}
\def\authorrunning{C. Y. Yin \& P. M. Kogge}
\begin{document}

\maketitle

\begin{abstract}
Subgraph Isomorphism uses a small graph as a pattern to identify within a larger graph a set of vertices that have matching edges. This paper addresses a logic program written in Prolog for a specific relatively complex graph pattern for which multiple conventional implementations (including parallel) exist. The goal is to understand the complexity differences between programming logically and programming conventionally. Discussion includes the process of converting the graph pattern into logic statements in Prolog, and the resulting characteristics as the size of the graph increased. The analysis shows that using a logic paradigm is an efficient way to attack complex graph problems.

\end{abstract}

\section{Introduction}\label{sec:intro}

The processing of graphs are of growing importance to many areas of computing. One problem of particular interest is ``subgraph isomorphism'' --- searching in a large graph for a set of vertices and edges that match some smaller pattern graph. 
This project utilizes the logic language Prolog to define rules that represent particular patterns and criteria within a graph, and then compares a variety of implementations, both logic and conventional. The logic implementations more than hold their own.

\section{Prior Work}\label{sec:prior}

Prolog \cite{prologprogramming} is a programming language based on Horn clause logic with a long history of use in tackling problems with complex patterns. It has three types of statements: a \textit{fact} is the name of a relation along with a tuple of argument values that is assumed to be part of the set defining the relation, and thus a ``true'' statement. A \textit{rule} has as its lefthand side a similar relation name and argument tuple that is assumed to be true if all the relations/tuples in a list on the right are provably true. A \textit{query} is a relation/tuple whose veracity is to be shown. 
The database of facts in Prolog makes it ideal to solve graph problems  \cite{Programmer-passport-prolog}. In Prolog, facts within the database can represent graph vertices, edges and properties, with the fact's arguments denoting two vertices connected by an edge. For example, the fact \texttt{edge(a,b)} represents an edge between vertices \texttt{a} and \texttt{b}. 

This project was inspired by the use of Prolog to encode the British Nationality Act \cite{nat-act}, where translating laws into Prolog rules allowed contradictions to be found.

\section{The Graph Problem}\label{sec:graph}

\begin{figure*}\begin{centering}
\includegraphics[width = 0.8\columnwidth]{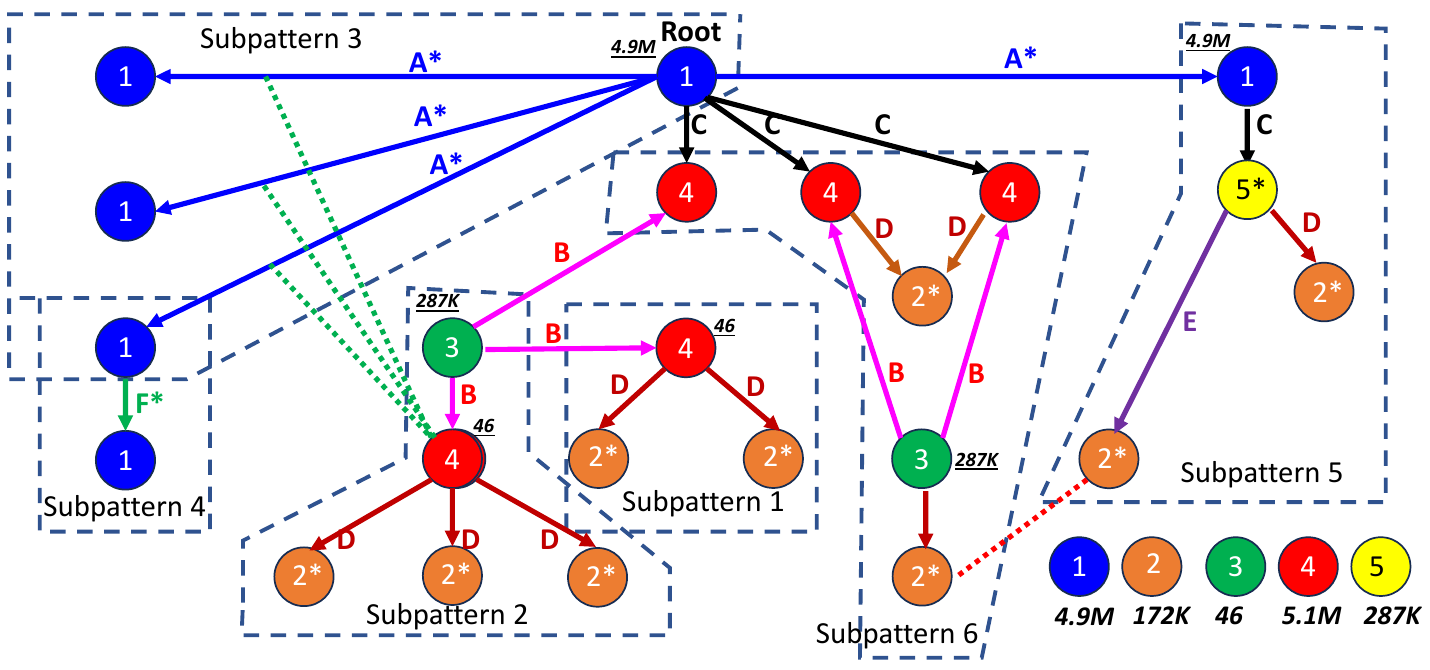}
\caption{Subgraph Pattern.}
\label{fig:subgraph}
\end{centering}\end{figure*}

The specific problem studied here originated from the U.S. IARPA AGILE program \cite{agile}, which provided reference implementation, data set, and timing.
It is a variant of a subgraph isomorphism where the target graph is built from five types of vertices and six types of directed edges, all of which have one or more properties associated with them. 
Fig. \ref{fig:subgraph} diagrams a simplified version of the subgraph pattern of interest \cite{agile-workflows}. Vertices of the same type are shown by circles of the same color, labeled ``1'' through ''5''. Edges are also color-coded, with labels ``A'' through ``F''; for simplicity, edges of type ``A'' are all edge type ``F'' reversed. An asterisk on a vertex or edge indicates that some property of the vertex or edge must have a specific value to satisfy the pattern \footnote{The formal description of the graph ontology has names for the vertex and edge classes, such as ``person'' for type 1 vertex. We use simple letters and numbers here to simplify the graph.}. The dotted green lines represent a separate relationship that must hold between a property of a type A edge and that of a type 4 vertex. The red dotted line also indicates a specific relationship must hold between the two vertices of type 2.
The key on the lower right gives the number of vertices for the reference dataset (about 2GB in size with about 38 million edges).
For each subgraph found in the graph that matches the subgraph pattern, the type 1 vertex labelled ``Root'' is to be returned. A graph generator was available to create graphs of different sizes. All graphs used in this paper have exactly one matching root vertex.

\section{Expression as a Logic Problem in Prolog}\label{sec:prolog}
To express the subgraph pattern as a logic problem in Prolog, we first converted the edges and vertices of the subgraph into a database of Prolog facts, with each fact defining the properties of a vertex or edge \cite{YinProlog}.
The fact predicates are named according to the 5 vertex and 6 edge types. Furthermore, the properties of each fact are represented as Object ID in the argument values of fact. For example, a type 1 vertex is represented as the fact:
\texttt{vertex1(735713441679521195)}.
This fact defines Object ID \texttt{735713441679521195} as a type 1 vertex. Similarly, edge type facts contain arguments specific to the edge. The following is the Prolog fact for a type D edge: \texttt{edgeD(932362105613871012, 60, 1)}. The extra arguments (``1'' here) correspond to specific D edge properties.

\subsection{Prolog Implementation Based on Subpatterns}

Prolog rules are used to express facts that depend on groups of other facts \cite {prologprogramming}. In our first Prolog implementation, we translated each of the 6 subgraph subpatterns into separate logical rules, and a single overarching rule uses all the subpattern rules to find and return the root vertex. The clauses in the rule body are joined by conjunctions, meaning a subpattern is true if and only if all clauses are true. The goal of each subpattern rule is to identify a vertex that holds true for the all the vertex and edge properties, allowing it to be used in identifying the Root vertex. 
In Prolog, ``variables'' start with an upper case letter.
For example, the following is the Prolog rule for subpattern 1: 
\begin{lstlisting}[style=myPrologstyle]
    subpattern1(Vertex4) :- edgeD(Vertex4, 69871376, 1), 
        edgeD(Vertex4, 1049632, 1), vertex4(_, Vertex4, _).
\end{lstlisting}

The rule states that \texttt{Vertex4} is found if the following conditions are met: a D edge connects a type 4 vertex to a type 2* vertex with the Object ID \texttt{69871376}, a second D edge connects the same type 4 vertex to a type 2* vertex with the Object ID \texttt{1049632}, and this type 4 vertex exists in the database\footnote{See \cite{YinProlog} for the source code of the program.}.
 
To find vertex 3, subpattern 1 and subpattern 2 are both queried in another rule. Both subpattern 1 and 2 rules look for vertices meeting the criteria of the D edges and type 2* vertices, thus identifying two type 4 vertices. With \texttt{Vertex4} identified in both subpatterns, the Object IDs of \texttt{Vertex4} are used to find a B edge that connects a type 3 vertex to both \texttt{Vertex4} vertices. The combined subpattern rule then identifies and returns the Object ID of vertex 3.

The rules for the other subpatterns follow a similar structure where the rule body specifies the properties to be met and the rule returns the Object ID of the vertex satisfying the criteria. The final query to find the Root vertex is an overarching rule that contains the rules for each subpattern.

\subsection{Prolog Implementation Using a Single Unified Rule}
A second Prolog implementation employs a singular rule to identify the Root vertex. Building on the subpattern implementation, this single rule unifies the subpatterns and maps the criteria needed to identify the Root vertex. The body of the rule consists of 35 clauses representing the vertices connected to the root node and the edges between the vertices. These clauses are joined by conjunctions, meaning the Root vertex is found if and only if all the clauses are true. The order of the clauses is grouped by subpatterns: subpattern 1, subpattern 2, subpattern 3, subpattern 4, subpattern 5, and subpattern 6. Unlike the subpattern program, this implementation directly passes the output values of predicates into subsequent clauses as arguments, avoiding repeated queries. The query for the single rule returns the Object ID of the Root vertex.
The pseudocode for the single rule is: 
\begin{verbatim}
    X is Root vertex if :-
        Vertex 3 is found from subpatterns 1 & 2,
        X is found from subpatterns 3 & 4,
        X has an A* edge to Vertex 1 found from subpattern 5, 
        X has three C edges to Vertex 4s found from subpattern 6, and
        Vertex 3 has a B edge to Vertex 4 found from subpattern 6. 
\end{verbatim}

\subsection{Handling Separate Relationships}
The subgraph pattern contains two separate unique relationships denoted by a dotted red line and dotted green lines. In the subpattern implementation, each of the two separate relationships has its own rule. These rules verify the existence of their respective relationships. Specifically, the rule for the red line checks for a particular relationship between the two type 2* vertices. The rule for the green line checks if there is a specific relationship between a type A* edge and a type 4 vertex. The rule for the green relationship is subsequently queried within the rule for subpattern 3, while the rule for the red relationship is queried within the rule for subpattern 6. The single rule implementation handles the separate red and green relationships in a manner similar to the subpattern program. However, the verification of these relationships is conducted within the single query rather than as separate rules.

\section{Prolog Execution Properties}\label{sec:logic}

\begin{figure}\begin{centering}
\includegraphics[width=\columnwidth]{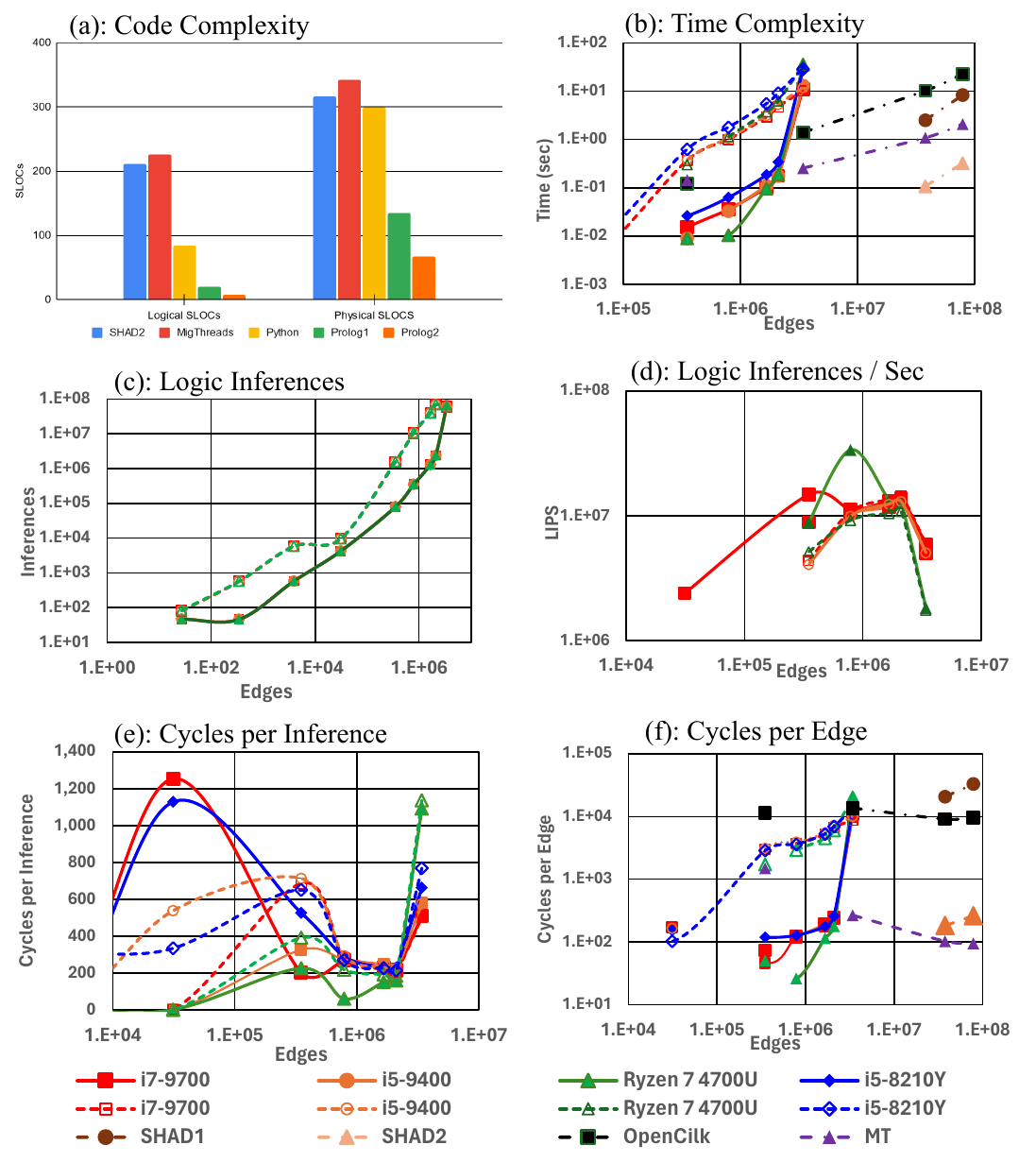}
\caption{Implementation Comparisons. Solid lines labelled with chip names are single core Prolog single pattern implementations. Short dotted lines are single core Prolog subpattern implementations. Long dashed lines are non-Prolog multi-core parallel implementations. }
\label{fig:comparisons}
\end{centering}\end{figure}

\begin{table}[t]
 \centering
 \caption{Hardware Platform Characteristics}\label{tab:systems}
 \small{
 \begin{tabular}{|c|c|c|c|c|c|c|c|}
   \hline\hline
   Prog.&&Avail.&Cores/&Core&Cores&L3&Memory\\
   Model&Processor&Nodes&Node&Clock&Used&Cache&Bandwidth \\
   \hline
   Prolog1-S,M&Intel i5-8210Y &1&2&1.6GHz&1&4MB&25.6GB/s \\ \hline 
   
   Prolog2-S,M&AMD 4700U &1&8&2.0GHz&1&8MB&51.2GB/s \\ \hline
   
   Prolog3-S,M&Intel i5-9400 &1&6&2.9 GHz&1&9MB&42.6GB/s \\ \hline
   
   Prolog4-S,M&Intel i7-9700 &1&8&3.0GHz&1&12MB/s&42.7GB/s \\ \hline
   Python&AMD 4700U&1&8&2.0GHz&1&18MB&51.2GB/s \\ \hline 
   SHAD-1&AMD EPYC 7763&16&128&2.45GHz&$\approx 24$&256MB&204GB/s \\ \hline 
   SHAD-2&E5-2680 v2&16&20&2.8GHz&20&25MB&51.2GB/s \\ \hline 
   Cilk&Xeon Silver 4208&1&16&2.1GHz&16&11MB&115.2GB/s \\ \hline 
   Mig. Thread&FPGA Custom&16&16&225MHz&16&0&34.1GB/s \\ 
   \hline\hline
    \end{tabular}
    }
\end{table}

We ran the two versions of Prolog code discussed above on four different systems with four different processors (listed as the first four rows of Table \ref{tab:systems}). In each case a single core was assumed. The first system used SWI-Prolog version 9.0.4; the others used version 9.3.7. 

In both Prolog implementations, we observed a phenomenon when repeating the same query. From the first to the second query, the number of inferences decreased by 18,657. After this initial decrease, the number of inferences remained constant for all following queries. This phenomenon was consistent across different datasets, Prolog versions, and hardware platforms; it is attributed to some sort of startup costs in the Prolog system, and was ignored in the following comparisons.

Fig. \ref{fig:comparisons}(b)-(e) lists some comparative properties of the Prolog runs against different sized graphs. Fig. \ref{fig:comparisons}(b) shows the execution time as a function of the number of edges in the graph. The solid lines are for the unified single query code, with a different color for each processor. The short-dashed lines of the same color are the execution times for the code which explicitly finds the subpatterns individually as separate goals\footnote{Ignore for now the lines to the right of about 4 million edges - these will be discussed in the next section.}. There are three key observations here. First, the subpattern code takes longer than the unified query until we get to graphs of about 4 million edges.

Second, the subpattern code is fairly linear in edge count up to at least 4 million edges\footnote{None of the systems would run problems bigger than about 4 million edges.}. Third, at around 2 million edges, something happens for the unified code to cause a major increase in the time. More than 4 million edges would not run.

One of the metrics returned by the Prolog system is the number of ``inferences'' performed. An inference occurs when all the goals on the righthand side of a rule have been satisfied, and the goal on the lefthand side is then asserted.
Fig. \ref{fig:comparisons}(c) lists the number of inferences made by each code in solving the problem. As seen in the figure, both codes grow the number of vertices at about the same rate, with the subpattern version typically requiring up to 10X more inferences. 

Fig. \ref{fig:comparisons}(d) uses these inference counts to compute an average ``logical inferences generated per second'' (LIPS). All systems seem to converge to around 10 million per second, with a die-off at around 2 million edges.

Finally, Fig. \ref{fig:comparisons}(e) tries to remove some of the discrepancy in underlying technology (clock rates among the various processors range up to about 2X) by computing the average number of clock cycles needed to compute the average inference.

As can be seen from the figure, the different processors and different implementations vary wildly until about 1 million edges, where they all converge at about 200 cycles per inference before exploding upwards at about 4 million edges. The cause of this increase is unclear. As seen from Table \ref{tab:systems}, there is a wide range in both last level cache and memory bandwidth, thus excluding memory access resources as a likely candidate. We speculate that this jump is caused by something in the SWI-PROLOG implementation.

\section{Comparison with Conventional Implementations}\label{sec:conventional}

This problem was chosen as a study target partially because there were multiple other implementations using a mix of both programming models and hardware architectures - none using any form of logic programming. The bottom 5 rows of Table \ref{tab:systems} summarize these combinations as follows:

\begin{itemize}[noitemsep,nolistsep,leftmargin=*]

\item \textit{Python}: a single-threaded implementation written in Python.

\item \textit{SHAD}: (\textit{Scalable High-Performance Algorithms and Data Structures}) \cite{SHAD-2018}  is a C++ library that provides a common shared-memory, task-based, programming model for a variety of hardware architectures ranging from multi-core shared memory systems to multi-node distributed memory systems. The same SHAD code was run on two different systems - one with a modern hardware base but an immature runtime (denoted SHAD-1), and the second on an older and smaller system with a significantly more mature software base (denoted SHAD-2).

\item \textit{Cilk}: an extension of C \cite{10.1145/209936.209958,programming-cilk} designed to provide dynamic multi-threading in a multi-core shared memory system. Its main extensions are two keywords:

\begin{itemize}[noitemsep,nolistsep,leftmargin=*]

\item A \textit{cilk\_spawn} prefix to a normal function call that indicates to the run-time that, if resources allow, the function call can be computed concurrently by  a thread independent of the calling thread.

\item A \textit{cilk\_sync} statement (no arguments needed) indicates that execution cannot proceed beyond this point until all child threads spawned in this block have completed. 

\end{itemize}

\item \textit{Migrating Threads}: a very Cilk-like parallel model where the underlying deeply multi-threaded hardware handles threads that must move from one physical node to another without any explicit software  \cite{kogge-refs:Dysart:2016:HSN:3018843.3018845}. This removes the need for software for both a multi-threaded runtime, and explicit software to recognize that data is ``not local''. Georgia Tech's CRNCH Center \footnote{\url{https://research.gatech.edu/crnchs-rogues-gallery-wants-bring-weirdest-hardware-campus}} hosts two such systems.
\end{itemize}

Execution of the last three was on parallel multi-node systems, with each node using multiple cores. For comparison, we used the execution times on just a single node (albeit with multiple cores). 

Fig. \ref{fig:comparisons}(a, b, f) provide  comparisons to the Prolog systems. Fig. \ref{fig:comparisons}(a) looks at the size of the code base using two metrics: physical lines of code (right-hand bar graph), and  logical statements (left-hand graph - akin to the number of semicolons). The remarkable conclusion from this is the dramatic reduction in code complexity for the Prolog implementation, especially the single query form.

Fig. \ref{fig:comparisons}(b) appends the measured execution time for these parallel implementations to the times from the Prolog implementations. Excluding whatever is causing the growth in Prolog time at the 4 million edge point, these parallel implementations seem to fall in line with the trends in the Prolog times. This is remarkable because these other implementations are ``single node'' but not ``single core'' implementations. Each has 16 or more cores, each individually at least the equal to the Prolog system cores. One would have expected much more improvement from the parallel systems.

Finally, while we could have used the inference counts from the Prolog systems to augment Fig. \ref{fig:comparisons}(e), a possibly more revealing metric is leveraging the concept of a ``Traversed Edge''\footnote{See the ``TEPS'' (Traversed Edges per Second) metric used in many graph benchmarks such as the Graph500.}. 
We chose to compare a ``Cycles per Edge'' metric, calculated as the product of the number of cores, core clock speed, and measured execution time, divided by the number of edges. This metric is graphed in Fig. \ref{fig:comparisons}(f) for all implementations. Again, ignoring what happens at 2 million edges in the Prolog implementations, those same Prolog implementations more than hold their own, especially the single query code.

\section{Conclusion}\label{sec:conclusion}
This paper presents two versions of a Prolog program designed to handle a complex graph pattern. 

Using a multiplicity of metrics, we compare those implementations with multiple non-Prolog implementations that use a variety of different programming models. The Prolog source codes are simpler and vastly smaller than the more conventional forms, but are performance comparable, especially considering they are not parallelized like the others.

Future work includes investigating the performance spike at 2 to 4 million edges, and using the built-in predicate \textit{asserta}, which allows the addition of new clauses to the beginning of a database \cite{prologprogramming}. This would reduce the number of repeated queries and decrease the number of inferences made, improving efficiency. In addition, parallel versions of Prolog may further enhance performance.

\nocite{*}
\bibliographystyle{eptcs}
\bibliography{bib}

\end{document}